\documentclass{amsart}
\def\t{\hbox}

\def\f{\frac}

\def\q{\quad}
\def\p{\varphi}

\def\k{\kappa}
\def\ep{\epsilon}
\def\a{\alpha}

\def\d{\delta}
\def\be{\begin{equation}}
\def\ee{\end{equation}}
\def\bea{\begin{eqnarray}}
\def\eea{\end{eqnarray}}
\def\ba{\begin{array}}
\def\ea{\end{array}}

\def\pr{\prime}

\theoremstyle{definition}

\theoremstyle{remark}

\numberwithin{equation}{section}

\newcommand{\abs}[1]{\lvert#1\rvert}

\newfont{\Bb}{msbm8 scaled\magstep{1}}
\newcommand{\rc}{\mbox{\Bb R}}

\usepackage{graphics}

\begin{document}

\title
[Stability Index Method]
{Inverse Scattering by the Stability Index Method
}

\author{Semion Gutman}
\address{Department of Mathematics\\ University of Oklahoma\\ Norman,
 OK 73019, USA}
\email{sgutman@ou.edu}

\author{Alexander G. RAMM 
}
\address{ Department of Mathematics\\ Kansas State University\\
Manhattan, Kansas 66506-2602, USA}
\email{ramm@math.ksu.edu}

\author{Werner Scheid}
\address{ Institut f\"ur Theoretische Physik der
Justus-Liebig-Universit\"at Giessen,
    Heinrich-Buff-Ring 16, D 35392, Giessen, Germany.}
    \email {werner.scheid@theo.physik.uni-giessen.de}


\begin{abstract} 
A novel numerical method for solving inverse scattering
problem with fixed-energy data is proposed. 
The method contains a new important concept: the stability index
of the inversion problem. This is a number, computed from the data,
which shows how stable the inversion is. If this index is small, then the
inversion provides a set of potentials which differ so little, that 
practically one can represent this set by one potential. If this
index is larger than some threshold, then practically one concludes
that with the given data the inversion is unstable and the potential
cannot be identified uniquely from the data.  
Inversion of the
fixed-energy phase shifts for several model potentials
is considered.
The results show practical efficiency of the proposed method.
The method is of general nature and is applicable to a very wide variety
of the inverse problems.
\end{abstract}

\thanks{ AGR thanks DAAD for support}
\maketitle



\section{Introduction}
Let $q(x),\, x\in \rc^3,$ be a real-valued potential with compact
support. Let $R>0$ be a number
 such that $q(x)=0$ for $\abs{x} > R$. We also
assume that $q\in L^2(B_R)\,,\ B_R=\{x:\abs{x}\leq R, x\in \rc^3\}$. 
Let $S^2$ be the unit sphere, and $\alpha \in S^2$. For
a given energy $k>0$ the scattering solution $\psi(x,\alpha)$ is defined
as the solution of

\begin{equation}
\Delta \psi+k^2\psi-q(x)\psi= 0\,,\quad x \in \rc^3 
\end{equation}
satisfying the following asymptotic condition at infinity:

\begin{equation}
\psi=\psi_0+v,\quad \psi_0:=e^{ik\alpha\cdot x}\,,\quad \alpha\in S^2\,,
\end{equation}

\begin{equation}
\lim_{r\rightarrow\infty}\int_{\abs{x}=r}\left| \frac{\partial v}{\partial
r}-ikv\right|^2 ds=0\,.
\end{equation}

It can be shown, that 

\begin{equation}
\psi(x,\alpha)=\psi_0+A(\alpha',\alpha,k)
\frac{e^{ikr}}{r}+o\left(\frac{1}{r}\right)\,,\;
\text{as}\ \ r\rightarrow\infty\,, \quad
\frac{x}{r}=\alpha '\, \quad r:=|x|.
\end{equation}

The function $A(\alpha',\alpha,k)$ is called 
the scattering amplitude, $\alpha$ and $\alpha'$
are the directions of the incident and scattered waves, and $k^2$
is the energy, see  \cite{n}, \cite{r3}.

For spherically symmetric scatterers $q(x)=q(r)$ the scattering
amplitude satisfies $A(\alpha',\alpha,k)=A(\alpha'\cdot\alpha,k)$. 
The converse is established in
\cite{r6}. Following \cite{rs}, the scattering amplitude
for $q=q(r)$ can be written as

\begin{equation}
A(\alpha',\alpha,k)=\sum^\infty_{l=0}\sum^l_{m=-
l}A_l(k)Y_{lm}(\alpha')\overline{Y_{lm}(\alpha)}\,,
\end{equation}

where $Y_{lm}$ are the spherical harmonics, normalized in
$L^2(S^2),$ and the bar denotes the
complex conjugate.

The fixed-energy phase shifts $-\pi<\delta_l\leq\pi$ 
($\delta_l=\delta(l,k)$,  $k>0$ is fixed) are related to $A_l(k)$ 
(see e.g., \cite{rs}) by the formula:

\begin{equation}
A_l(k)=\frac{4\pi}{k}e^{i\delta_l}\sin(\delta_l)\,.
\end{equation}
In section 2 we give, following [1], formulas for calculating
fixed-energy phase shifts for 
piecewise-constant compactly supported
potentials. let us denote this class of potentials by PC.
Since an arbitrary integrable potential 
can be approximated
with the prescribed accuracy by a PC potential, the
class PC is sufficiently large for practical purposes.

In sections 3 and 4 a novel minimization method, 
the stability index method, is described.
Our inversion procedure is based on this method.
An important novel feature of this method,
which seems not have been present in other methods,
is the concept of the stability index, which is a 
number characterizing the stability of the numerical inversion.

Several parameter-fitting procedures were proposed for calculating the
potentials from the fixed-energy phase shifts, (by Fiedeldey,
Lipperheide, Hooshyar and Razavy, Ioannides and Mackintosh, Newton,
 Sabatier, May and Scheid, Ramm, and others). 
These works  are referenced and their results are described in [5]
and [10]. Recent works [6]-[9] and [18]-[20] present
new numerical methods for solving this problem.

In section 5 numerical examples of the inversion
of the fixed-energy shifts are given for three potentials.
Physical motivation for the choice of these potentials 
is given and directions for future research are suggested.

Section 6 contains brief conclusions.

\section{Phase Shifts for Piecewise-Constant Potentials}

Phase shifts for a spherically symmetric potential can be
computed by a variety of methods, e.g. by a variable phase method
described in \cite{calogero}. The computation involves solving
a nonlinear ODE for each phase shift. However, if the potential is compactly
supported and piecewise-constant, then
a much simpler method described in \cite{ars} can be used.
It is summarized below. Since the set of compactly
supported and piecewise-constant potentials is dense in the set
$L^1(0, \infty)$ potentials, it is quite reasonable to look for an
approximate solution to the inverse scattering problem in the class
of piecewise-constant compactly supported potentials. 

Consider a finite set of points $0=r_0< r_1 <r_2<\dots<r_N=R$
and a piecewise-constant potential
\be\label{pot}
q(r)=q_i,\t{ on } [r_{i-1},r_i) \t{ for } 
i=1,\dots, N, \t{ and } q=0\t{ for }r\ge R.
\end{equation}

Denote 
\be\label{phase2}
\k_i^2:=k^2-q_i\,,
\end{equation}
where $i=1,\dots, N,$
and $k$ is some fixed positive number.
Consider the following problem for the radial Schr\"odinger equation:
\be
\f{d^2\p_l}{dr^2}+\Biggl(k^2-\f{l(l+1)}{r^2}\Biggl)\p_l=q\p_l,
\q \lim_{r\to 0}(2l+1)!!r^{-l-1}\p_l(r)=1,
\end{equation}

which we rewrite as:

\be\label{sc}
\f{d^2\p_l}{dr^2}+\Biggl(\k_i^2-\f{l(l+1)}{r^2}\Biggl)\p_l=0
\end{equation}

on the interval $r_{i-1}\le r < r_i$.
On $[r_{i-1},r_i)$ one has the following general solution of (\ref{sc})

\be\label{phase5}
\p_l(r)=A_ij_l(\k_ir)+B_in_l(\k_ir),
\end{equation}
where
\be
j_l(kr)=\sqrt{\frac{\pi kr}2}J_{l+1/2}(kr)\,,\;
n_l(kr)=\sqrt{\frac{\pi kr}2}N_{l+1/2}(kr)\,
\end{equation}
and $J_l\,,\ N_l$ are the Bessel and Neumann functions.

If $\k_i^2=k^2-q_i \leq 0$
for some $i$, then the solution $\p_l(r)$ in (\ref{sc}) can be expressed
through some powers of $r$ (for $\k_i=0$) or the modified Bessel and
Neumann functions,
and our approach is still valid with the appropriate changes.

From the regularity of $\p_l$ at zero one gets $B_1=0$. Denote
$x_i=B_i/A_i$, then $x_1=0$. 
We are looking for a continuously differentiable solution $\p_l$.
Thus, the following interface conditions hold:
\be\ba{lcc}
A_ij_l(\k_ir_i)+B_in_l(\k_ir_i)=A_{i+1}j_l(\k_{i+1}r_{i})+
B_{i+1}n_l(\k_{i+1}r_{i}),\\ \\
\f{\k_i}{\k_{i+1}}[A_ij_l^\pr(\k_ir_i)+B_in_l^\pr(\k_ir_i)]=
A_{i+1}j_l^\pr(\k_{i+1}r_{i})+B_{i+1}n^\pr_l(\k_{i+1}r_{i}).
\end{array}
\end{equation}

Therefore
\be
\begin{pmatrix}
A_{i+1}\\ B_{i+1}
\end{pmatrix}
=\f{1}{\k_{i+1}}
\begin{pmatrix}\a^i_{11} & \a^i_{12}\cr
\a^i_{21} & \a^i_{22}
\end{pmatrix}
\begin{pmatrix}A_{i}\cr B_{i}
\end{pmatrix},
\end{equation}
where the entries of the matrix $\alpha^i$ can be written explicitly (see \cite{ars}
for details).

Thus
\be\label{xk}
x_{i+1}=\f{\a^i_{21}+\a^i_{22}x_i}{\a^i_{11}+\a^i_{12}x_i},\q
x_i:=\f{B_i}{A_i}
\end{equation}

The phase shift $\d(k,l)$ is defined by

\be
\p_l(r)\sim{|F(k,l)|\over k^{l+1}}
\sin(kr-\frac{\pi l}{2}+
\delta(k,l))\quad r\to\infty\enspace ,
\end{equation}

where $F(k,l)$ is the Jost function.
For $r>R$ one has:

\be\label{as}
\p_l(r)=A_{N+1}j_l(kr)+B_{N+1}n_l(kr).
\end{equation}

From (\ref{as}) and the asymptotics
$j_l(kr)\sim\sin(kr-l\pi/2),\q n_l(kr)\sim-\cos(kr-l\pi/2)$,
$r\to\infty$, one gets:

\be\label{dkl}
\tan\delta(k,l)=-\f{B_{N+1}}{A_{N+1}}=-x_{N+1}\,.
\end{equation}
Finally, the phase shifts of the potential $q(r)$
are calculated by the formula:
\be\label{dklf}
\delta(k,l)=-\arctan x_{N+1}.
\end{equation}

Let $q_0(r)$ be a spherically symmetric piecewise-constant potential.
Let $\{\tilde\delta(k,l)\}_{l=1}^N$ be the set of its phase shifts
for a fixed $k>0$ and a sufficiently large $N$.
Let $q(r)$ be another  potential, and let
 $\{\delta(k,l)\}_{l=1}^N$ be the set of its phase shifts.

The best fit to data
function $\Phi(q,k)$ is defined by

\begin{equation}\label{phi}
\Phi(q,k)=\frac{\sum^N_{l=1}\abs{\delta(k,l)-
\tilde\delta(k,l)}^2}
{\sum^N_{l=1}\abs{\tilde\delta(k,l)}^2}.
\end{equation}

The phase shifts are known to decay rapidly with $l$, see \cite{rai}.
Thus, for sufficiently large $N$, the function $\Phi$ is practically the same as
the one which would use all the shifts in (\ref{phi}).
The inverse problem of the reconstruction of the potential from its
fixed-energy phase shifts is reduced to the minimization of the 
objective function $\Phi$
over an appropriate admissible set.

\section{Stability Index Minimization Method}

Let the minimization problem be

\begin{equation}\label{sta1}
\min\{\Phi(q) \ : \ q\in A_{adm}\}
\end{equation}

Let $\tilde q_0$ be its global minimizer. Typically, the
structure of the objective function $\Phi$ is quite complicated: this
function may have
many local minima. Moreover, the objective function in a neighborhood of
 minima can be nearly
flat resulting in large minimizing sets defined by

\begin{equation}\label{sta2}
S_{\ep}=\{q\in A_{adm}\ :\ \Phi(q)<\Phi(\tilde q_0)+\ep \}
\end{equation}
for an $\ep>0$.

Given an $\ep>0$, let $D_\ep$ be the diameter of the minimizing set $S_\ep$, which we call
the {\bf Stability Index $D$} of the minimization problem
(\ref{sta1}). The usage of the letter $D$ for this index is
explained in formula (3.6) below. 

One would expect to obtain stable identification for minimization
problems with small stability indices. However, the minimization
problems with large stability indices have distinct minimizers with
practically the same values of the objective function. If no additional
information is known, one has an uncertainty
of the minimizer's choice. The stability index provides a quantitative
measure of this uncertainty or instability of the minimization.

The basic idea of the Stability Index minimization method is to iteratively estimate
normalized stability indices of a minimization problem, and, based on
this information, to conclude if the method has achieved a stable minimum.

A particular implementation of the Stability Index method used here
employs a Hybrid Stochastic-Deterministic (HSD) approach. The stochastic
part explores the entire admissible set, while the deterministic local
minimization finds the best fit in a neighborhood of the chosen 
in the stochastic part of the search initial
guesses. The HSD approach has proved to be successful for a variety of
problems 
in inverse quantum scattering (see \cite{gut5,gutmanramm2}) as well as in other
applications (see \cite{gutmanramm,gut3}). A somewhat different
implementation of the Stability Index Method is described in
\cite{r423}.

We seek the potentials $q(r)$ in the class of piecewise-constant, spherically 
symmetric real-valued functions. Let the admissible set be
\begin{equation}\label{adm}
A_{adm} \subset \{(r_1,r_2,\dots,r_M,q_1,q_2,\dots,q_M)\ : \ 0\leq r_i\leq R\,,\ 
q_{low}\leq q_m \leq q_{high}\}\,,
\end{equation}

where the bounds $q_{low}$ and $q_{high}$
for the potentials, as well as the bound $M$ on
the expected number of layers are assumed to be known.

A configuration $(r_1,r_2,\dots,r_M,q_1,q_2,\dots,q_M)$ corresponds to
the potential

\begin{equation}
q(r)=q_m\,,\quad \text{for}\quad r_{m-1}\leq r<r_m\,,\quad 1\leq m\leq M\,, 
\end{equation}
where $r_0=0$ and $q(r)=0$ for $r\geq r_M=R$.

Note, that the admissible configurations must also satisfy

\begin{equation}\label{admr}
r_1\leq r_2\leq r_3 \leq\dots\leq r_M\,.
\end{equation}

First we describe the global (stochastic) part of the algorithm, which
can be called the Iterative Reduced Random Search (IRRS) method. 
This description is followed by
its iterative version, and a Local Minimization Method (LMM)
incorporating a Reduction procedure.

Let a batch $H$ of $L$ trial points be generated in
$A_{adm}$ using a uniformly distributed random variable.
In our case $A_{adm}$ is a box in $\rc^{2M}$. The uniform random variable is
called $2M$ times to produce a point (configuration representing a
potential) in this box (after the appropriate rescaling
 in each dimension). Finally, the obtained values of $r_i$ are rearranged in 
the ascending order to satisfy (\ref{admr}).

A certain fixed fraction  $\gamma$ of the original batch of $L$
points is used to proceed with the local searches. Typically, $L=5000$ and $\gamma=0.01$.
This reduced sample $H_{red}$ of
$\gamma L$ points  is chosen to contain the points with the smallest
$\gamma L$ values of $\Phi$ among the original batch $H$. The local
searches (the LMM procedure)
are started from every point in this reduced sample $H_{red}$ . This way only the points that seem to
be in a neighborhood of the global minimum are used for an expensive local
minimization, and the computational time is not wasted on less
promising candidates.

 Let $H_{min}$ be the 
$\gamma L$ points obtained as the result of the local minimizations 
($\gamma L=50$ in our computations). Let
$S_{min}$ be the subset of $H_{min}$ containing points
 $\{p_i\}$ with the smallest 
$\nu\gamma L$ ($0<\nu<1$, we used $\nu=0.16$) values in $H_{min}$. We call $S_{min}$ the
minimizing set. The choice of $\nu$ determines a representative sample
of global minimizers. If all these minimizers are close to each other,
then  the objective function $\Phi$ is not
flat near the global minimum. That is, the method identifies the minimum consistently. 
 To define this consistency in quantitative terms, let
$\|.\|$ be a norm in the admissible set.

Let
\begin{equation}\label{diam}
D=diam(S_{min})=\max\{\|p_i-p_j\|/d_{av}\ :\ p_i,p_j\in S_{min}\}\,,
\end{equation}
where $d_{av}$ is the average norm of the elements in $H_{min}$.
The normalization by $d_{av}$ is introduced to provide comparable results for different potentials.
Thus $D$ is an estimate for the (normalized) {\bf Stability Index} of
the minimization problem. The identification is considered to be stable
if the Stability Index $D<\epsilon$. Otherwise, another batch of trial
points is generated, and the process is repeated as follows.

\subsection*{Iterative Reduced Random Search (IRRS)} (at the $j-$th 
iteration).

Fix $0<\gamma, \nu <1,\ \beta > 1,\ \epsilon>0$ and $j_{max}$.

\begin{enumerate}
\item  Generate another batch $H^j$ of $L$ trial points (configurations) in $A_{adm}$ using a uniform
random distribution.  

\item  Reduce $H^j$ to the reduced sample $H^j_{red}$ of $\gamma L$ points by selecting 
the points in $H^j$ with the smallest $\gamma L$ values of $\Phi$.

\item  Apply the Local Minimization Method (LMM) starting it at each of the $\gamma L$
points in $H^j_{red}$,
 and obtain the set $H^j_{min}$ consisting of the $\gamma L$ minimizers.

\item Combine $H^j_{min}$ with $H^{j-1}_{min}$ obtained at the previous
iteration. Let $S^j_{min}$ be the set of $\nu\gamma L$
points from $H^j_{min}\cup H^{j-1}_{min}$ with the smallest
 values of $\Phi$. (Use $H^1_{min}$ for $j=1$).

\item  Compute the Stability Index (diameter) $D^j$ of $S^j_{min}$ by 
$D^j=\max\{\|p_i-p_k\|/d_{av}\ :\ p_i,p_k\in S_{min}\}\,,$
where $d_{av}=\sum \|p_k\|/\gamma L\,,\ p_k\in H^1_{min}$ is the average norm of the elements 
of $H^1_{min}$.
(Thus, $d_{av}$ is computed only once
at $j=1$ and this value is used for all subsequent iterations).

\item Stopping criterion. 

Let $p\in S^j_{min}$ be the point with the smallest value
of $\Phi$ in $S^j_{min}$ (the global minimizer).

If $D^j\leq\epsilon$, then stop. The global minimum is $p$. The
minimization is stable.

If $D^j>\epsilon$ and $\Phi(q)\leq\beta \Phi(p)\ : \ q\in S^j_{min}$, then  stop.
The minimization is unstable. The Stability Index $D^j$ is the measure of the
instability of the minimization.

Otherwise, return to step 1
and do another iteration, unless the maximum number of iterations $j_{max}$ is
exceeded.

\end{enumerate}

We used $\beta=1.1$, $\epsilon=0.02$ and $j_{max}=30$. The choice of
these and other parameters 
 ($L=5000,\, \gamma=0.01,\ \nu=0.16\,\ \epsilon_r=0.1$ (used in LMM))
 is dictated by their meaning in the algorithm and the comparative performance
of the program at their different values. As usual, 
some adjustment of the parameters, stopping criteria,
etc., is needed to achieve the optimal performance of the algorithm.

\section{Local Minimization Method}

The Hybrid Stochastic Deterministic Method couples the Stochastic part described in the previous section
with a deterministic Local Minimization Method.
Numerical experiments show that the
objective function $\Phi$ is relatively well behaved in this problem:
while it contains many local minima and, at some points, $\Phi$ is not
differentiable, standard minimization methods work well here. 
A Newton-type method for the minimization of $\Phi$ is described in
\cite{ars}. We have chosen to use a variation of Powell's minimization
method which does not require the computation of the derivatives of the
objective function. Such method needs a minimization routine for 
a one-dimensional minimization of $\Phi$, which we do using a Bisection or a
Golden Rule method. See \cite{gut3} or \cite{gut5} for a complete
description of our method. 

Now we can describe our Basic Local Minimization
Method in $\rc^{2M}$, which is a modification 
of Powell's minimization method \cite{bre}. It is assumed here that the
starting position (configuration) $Q_0\in A_{adm}$ is suppied by the
procedure LMM (see below), and
the entry to LMM is provided by the global minimization part (IRRS).

\subsection*{Basic Local Minimization Method}
\begin{enumerate}

\item  Choose the set of directions $u_i\,,\;i=1,2,\dots,2M,$ to be the 
standard basis
 in $\rc^{2M}$
\[
u_i=(0,0,\dots,1,\dots,0)\,,
\]
where $1$ is in the i-th place.
\item  Save your starting configuration supplied by LMM as $Q_0$ .
\item  For each $i=1,\dots,2M$ move from $Q_0$ along the line defined by $u_i$ 
 and find the point of minimum $Q_i^t$. This defines $2M$ temporary
points of minima.
\item  Re-index the directions $u_i$, 
so that (for the new indices) $\Phi(Q_1^t)\leq \Phi(Q_2^t)
\leq,\dots,\Phi(Q_{2M}^t)\leq\Phi(Q_0)$.

\item  For $i=1,\dots,2M$ move from 
$Q_{i-1}$ along the direction $u_i$ and find the point
of minimum $Q_i$.
\item  Set $v=Q_{2M}-Q_0$.
\item  Move from $Q_0$  along the direction $v$ and find the minimum. Call it
 $Q_0$ again. It replaces $Q_0$ from step 2.
\item Repeat the above steps until a stopping criterion is satisfied.
\end{enumerate}

Note, that we use the temporary points of minima $Q_i^t$ only to rearrange the
initial directions $u_i$ in a different order. The stopping criterion is
the same as the one in \cite[Subroutine Powell]{numrec}.

Still another refinement of the local phase is necessary
to produce a successful minimization. The admissible set $A_{adm}$, see
(\ref{adm})-(\ref{admr}), belongs to a $2M$ 
dimensional minimization space $\rc^{2M}$.
The dimension $2M$ of this space
is chosen a priori to be larger than $2N$, where $N$
is the number of layers in the original potential. We have chosen
$M=2$ in our numerical experiments. 
However, since the sought potential may have fewer than 
$M$ layers, we found that conducting searches in lower-dimensional subspaces
of $\rc^{2M}$ is essential for the local minimization phase.
A variation of the following "reduction" procedure has also been found 
to be necessary in \cite{gutmanramm} for the search of small subsurface objects, 
and in \cite{gut3} for the 
identification of multilayered scatterers.

If two adjacent layers in a potential
have  values $v_{i-1}$ and $v_i$ and the objective function $\Phi$
is not changed much when both layers are 
assigned the same  value $v_i$ (or $v_{i-1}$),
 then these two layers can be replaced with just one layer occupying their
place. The change in $\Phi$ is controlled by the parameter $\epsilon_r$.
We used $\epsilon_r=0.1$. This value, found from numerical
experiments, seems to provide the most consistent identification.
The minimization problem becomes
constrained to a lower dimensional subspace of $\rc^{2M}$ and the local
minimization is done in this subspace.

\subsection*{Reduction Procedure}

Let $\epsilon_r$ be a positive number.

\begin{enumerate}

\item  Save your starting configuration
$Q_0=(r_1,r_2,\dots,r_M,v_1,v_2,\dots,v_M)\in A_{adm}$
 and the value $\Phi(Q_0)$. Let the $(M+1)$-st 
layer be $L_{M+1}=\{r_M\leq |x| \leq
R\}$ and $v_{M+1}=0$.

\item  Let $2\leq i\leq M+1$. Replace $v_{i-1}$ 
in the layer $L_{i-1}$ by $v_i$. This defines a new configuration $Q_i^d$,
where the layers $L_{i-1}$ and $L_i$ are replaced with one new layer. Here $d$ stands
for the downward adjustment. Compute
$\Phi(Q_i^d)$ and the difference
$c_i^d=|\Phi(Q_0)-\Phi(Q_i^d)|$. Repeat for each layer in the original
configuration $Q_0$.

\item  Let $1\leq i\leq M$. Replace $v_{i+1}$ 
in the layer $L_{i+1}$ by $v_i$.  This defines a new configuration $Q_i^u$,
where the layers $L_i$ and $L_{i+1}$ are replaced with one new layer. Here $u$ stands
for the upward adjustment. Compute
$\Phi(Q_i^u)$ and the difference
$c_i^u=|\Phi(Q_0)-\Phi(Q_i^u)|$. Repeat for each layer in the original
configuration $Q_0$.

\item  Find the smallest among the numbers $c_i^d$ and $c_i^u$.
If this number is less than $\epsilon_r\Phi(Q_0)$, then implement the adjustment that
produced this number. The resulting new configuration has one less layer
than the original configuration $Q_0$.

\item Repeat the above steps until no further reduction in the number of
layers is occurring.

\end{enumerate}

Note, that an application of the Reduction Procedure may or may not
result in the actual reduction of the number of layers.

Finally, the entire Local Minimization Method {\bf (LMM)} consists of the
following:

\subsection*{Local Minimization Method (LMM)}
\begin{enumerate}

\item  Let your starting configuration supplied by IRRS be
$Q_0=(r_1,r_2,\dots,r_M,v_1,v_2,\dots,v_M)\in A_{adm}$.

\item  Apply the Reduction Procedure to $Q_0$, 
and obtain a reduced configuration
$Q_0^r$ containing $M^r$ layers.

\item  Apply the Basic Minimization Method in $A_{adm}\bigcap \rc^{2M^r}$ 
with the starting configuration $Q_0^r$, and obtain a configuration $Q_1$.

\item  Apply the Reduction Procedure to 
$Q_1$, and obtain a final reduced configuration
$Q_1^r$.

\end{enumerate}

As we have already mentioned, LMM is used as the local phase of the
global minimization.

\section{Numerical Results}
We studied the performance of the algorithm for 3 
different potentials $q_i(r),\ i=1,2,3$ chosen from the physical
considerations.

The potential $q_3(r) = - 10$ for $0\leq r < 8.0$ and $q_3 = 0 $
for $r\geq 8.0$ and a wave number $k=1$ constitute a typical
example  for elastic scattering of neutral particles
in nuclear and atomic physics. In nuclear physics one measures the length in
units of fm = $10^{-15}$m, the quantity $q_3$ in units of
1/fm$^2$, and the wave number in units of 1/fm. The physical
potential and incident energy are given by $V(r) = \frac
{\hbar^2}{2\mu} q_3(r)$ and $E = \frac {\hbar^2 k^2}{2\mu}$,
respectively. 
here $\hbar:= \frac {h}{2\pi}$,  $h=6.625 10^{-27}$ erg$\cdot$s 
is the Planck constant, 
 $\hbar c=197.32$  MeV$\cdot$fm, $c=3\cdot 10^6$ m/sec is the
velocity of light, 
and $\mu$ is the mass of a neutron.
By choosing the mass $\mu$ to be equal to
the mass of a neutron $\mu$ = 939.6 MeV/$c^2$, the potential and
energy have the values of $V(r)$ = -207.2 MeV for $0\leq r < 8.0$ fm
and $E( k=$1/fm ) = 20.72 MeV. In atomic physics one uses atomic
units with the Bohr radius $a_0 = 0.529\cdot10^{-10}$m as the unit of
length. Here, $r, k $ and $q_3$ are measured in units of $a_0,
1/a_0$ and $1/a_0^2$, respectively. By assuming a scattering of an
electron with mass $m_0$ = 0.511 MeV/$c^2$, we obtain the potential
and energy as follows: $V(r) = -136$ eV for $0\leq r < 8 a_0 =
4.23\cdot10^{-10}$m and $E( k=1/a_0 ) = 13.6$ eV. These numbers
give motivation for the choice of examples applicable in
nuclear and atomic physics.

The method used in this paper deals with finite-range
(compactly supported) potentials. One can use this method for
potentials with the Coulomb tail or other potentials
of interest in physics, which are not of finite range.
This is done by using the phase shifts transformation
method which allows one to transform the phase shifts corresponding to
a potential, not of finite range, whose behavior is known for
$r>a$, where $a$ is some radius, into the phase shifts corresponding
to a potential of finite range $a$ (see [2], p.156).

In practice differential cross section is measured 
at various angles, and from it the fixed-energy phase
 shifts are calculated by a parameter-fitting procedure.
Therefore, we plan in the future work to generalize
the stability index method to the
case when the original data are the values
of the differential cross section, rather than
the phase shifts.



By the physical reasons discussed above,
we choose the following three potentials:
\[
q_1(r)=\begin{cases} 
-2/3 & 0\leq r < 8.0\\
0.0 &  r \geq 8.0
\end{cases}
\]

\[
q_2(r)=\begin{cases} 
-4.0 & 0\leq r < 8.0\\
0.0 &  r \geq 8.0
\end{cases}
\]

\[
q_3(r)=\begin{cases} 
-10.0 & 0\leq r < 8.0\\
0.0 &  r \geq 8.0
\end{cases}
\]

In each case the following values of the parameters have been used. The
radius $R$ of the support of each $q_i$ was chosen to be $R=10.0$.
The admissible set $A_{adm}$ (\ref{adm}) was defined with $M=2$. The
Reduced Random Search parameters:
$L=5000\,,\;\gamma=0.01\,,\;\nu=0.16\,,\; \epsilon=0.02\,,\;\beta=1.10\,,j_{max}=30$.
The value $\epsilon_r=0.1$ was used in
the Reduction Procedure during the local minimization phase.
The initial configurations were generated using a random 
number generator with seeds determined by the system time.
A typical run time was about 10 minutes on a 333 MHz PC, 
depending on the number of iterations in IRRS.
The number $N$ of the shifts used in (\ref{phi}) for the formation of
the objective function $\Phi(q)$ was $31$ for all the wave numbers.
As it can be seen from Table 1 the shifts for the potential $q_3$ decay rapidly for $k=1$,
but they remain large for $k=4$.
 The upper and lower bounds for the potentials $q_{low}=-20.0$ and $q_{high}=0.0$
used in the definition of the admissible set $A_{adm}$ were chosen to
reflect a priori information about the potentials.



\begin{table}
\caption{Phase shifts $\delta(k,l)$ of $q_3(r)$.}

\begin{tabular}{r r r}

\hline

$l$ & $k=1.0$ & $k=4.0$ \\
\hline
  0  &  -0.66496 & -0.62217  \\
  1  &  -0.31009 & -0.64598  \\
  2  &  -0.72324 &  -0.65824 \\
  3  &  -0.88586 &  -0.64604 \\
  4  &  -0.74713 &  -0.74239 \\
  5  &   1.15839 &  -0.65260 \\
  6  &   1.54292 &  -0.86826 \\
  7  &  -0.56945 &  -0.69851 \\
  8  &  -0.38745 &  -1.00663 \\
  9  &   0.16888 &  -0.83757 \\
 10  &  -0.02690 &  -1.08641 \\
 11  &  -0.01261 &  -1.09074 \\
 12  &   0.00017 &  -1.08645 \\
 13  &  -0.00010 &  -1.39603 \\
 14  &  -0.00004 &  -1.24536 \\
 15  &   0.00000 &  -1.55399 \\
 16  &   0.00000 &  1.49036 \\
 17  &   0.00000 &  1.56437 \\
 18  &   0.00000 &  1.11836 \\
 19  &   0.00000 &  1.12265 \\
 20  &   0.00000 &  1.11829 \\
 21  &   0.00000 &  0.60125 \\
 22  &   0.00000 &  0.58327 \\
 23  &   0.00000 &  0.59973 \\
 24  &   0.00000 &  0.00875 \\
 25  &   0.00000 &  -0.18826 \\
 26  &   0.00000 & -0.11221  \\
 27  &   0.00000 &  -0.47503 \\
 28  &   0.00000 &  -1.22725 \\
 29  &   0.00000 &  -1.29222 \\
 30  &   0.00000 &  -1.25626 \\

\hline
\end{tabular}

\end{table} 

The identification was attempted with 3 different noise levels $h$.
The levels are $h=0.00$ (no noise), $h=0.01$ and $h=0.1$.
 More precisely, the noisy phase shifts $\delta_h(k,l)$ were obtained from
the exact phase shifts $\delta(k,l)$ by the formula

\[
\delta_h(k,l)=\delta(k,l)(1+(0.5-z)\cdot h)\,,
\]
where 
$z$ is the uniformly distributed on $[0,1]$ random variable.

The distance $d(p_1(r),p_2(r))$ for potentials in step 5 of the IRRS algorithm was
computed as 

\[
 d(p_1(r),p_2(r))=\|p_1(r)-p_2(r)\|\,
\]
where the norm is the $L_2$-norm in $\rc^3$.

The results of the identification algorithm (the Stability Indices) for
different iterations of the IRRS algorithm
are shown in Tables 2-4.


\begin{table}
\caption{Stability Indices for $q_1(r)$ identification at different noise levels $h$.}

\begin{tabular}{r r r r r}
\hline

$k$ & $Iteration$ & $h=0.00$ & $h=0.01$ & $h=0.10$\\

\hline

\hline

1.00 & 1 & 1.256985 & 0.592597	 & 1.953778	 \\
   & 2 &    0.538440 & 0.133685 & 0.799142 \\
   & 3 &    0.538253 & 0.007360 & 0.596742 \\
   & 4 &    0.014616 & & 0.123247 \\
 &5 &&& 0.015899 \\
\hline
2.00 & 1 & 0.000000 & 0.020204	 & 0.009607	 \\

\hline
2.50 & 1 & 0.000000 & 0.014553	 & 0.046275	 \\
\hline
3.00 & 1 & 0.000000 & 0.000501	 & 0.096444	 \\
\hline
4.00 & 1 & 0.000000 & 0.022935  & 0.027214	 \\
\hline

\hline
\end{tabular}
\end{table} 


\begin{table}
\caption{Stability Indices for $q_2(r)$ identification at different noise levels $h$.}

\begin{tabular}{r r r r r}
\hline

$k$ & $Iteration$ & $h=0.00$ & $h=0.01$ & $h=0.10$\\

\hline

\hline

1.00 & 1 & 0.774376 & 0.598471	 & 0.108902 \\
   & 2 &    0.773718 & 1.027345 & 0.023206 \\
   & 3 &    0.026492 & 0.025593 & 0.023206 \\
   & 4 &    0.020522 & 0.029533 & 0.024081 \\
   & 5 &    0.020524 & 0.029533 & 0.024081 \\
   & 6 &    0.000745 & 0.029533 &  \\
   & 7 &  & 0.029533 &  \\
   & 8 &  & 0.029533 &  \\
   & 9 &  & 0.029533 &  \\
   & 10 &  & 0.029533 &  \\
   & 11 &  & 0.029619 &  \\
   & 12 &  & 0.025816 &  \\
   & 13 &  & 0.025816 &  \\
   & 14 &  & 0.008901 &  \\

\hline
2.00 & 1 & 0.863796 & 0.799356	 & 0.981239 \\
 & 2 & 0.861842 & 0.799356 & 0.029445 \\
 & 3 & 0.008653 & 0.000993 & 0.029445 \\
 & 4 &  &  & 0.029445 \\
 & 5 &  &  & 0.026513 \\
 & 6 &  &  & 0.026513 \\
 & 7 &  &  & 0.024881 \\

\hline

2.50 & 1 &  1.848910 & 1.632298	 & 0.894087 \\
 & 2 & 1.197131 & 1.632298 & 0.507953 \\
 & 3 & 0.580361 & 1.183455 & 0.025454 \\
 & 4 & 0.030516 & 0.528979 & \\
 & 5 & 0.016195 & 0.032661 & \\

\hline
3.00 & 1 & 1.844702 & 1.849016	 & 1.708201 \\
 & 2 & 1.649700 & 1.782775	 & 1.512821 \\
 & 3 & 1.456026 & 1.782775	 & 1.412345 \\
 & 4 & 1.410253 & 1.457020	 & 1.156964 \\
 & 5 & 0.624358 & 0.961263	 & 1.156964 \\
 & 6 & 0.692080 & 0.961263	 & 0.902681 \\
 & 7 & 0.692080 & 0.961263	 & 0.902681 \\
 & 8 & 0.345804 & 0.291611	 & 0.902474 \\
 & 9 & 0.345804 & 0.286390	 & 0.159221 \\
 & 10 & 0.345804 & 0.260693	 & 0.154829 \\
 & 11 & 0.043845 & 0.260693	 & 0.154829 \\
 & 12 & 0.043845 & 0.260693	 & 0.135537 \\
 & 13 & 0.043845 & 0.260693	 & 0.135537 \\
 & 14 & 0.043845 & 0.260693	 & 0.135537 \\
 & 15 & 0.042080 & 0.157024	 & 0.107548 \\
 & 16 & 0.042080 & 0.157024	 &  \\
 & 17 & 0.042080 & 0.157024	 &  \\
 & 18 & 0.000429 & 0.157024	 &  \\
 & 19 & & 0.022988 &  \\

\hline
4.00 & 1 & 0.000000 & 0.000674  & 0.050705 \\
\hline

\end{tabular}
\end{table} 


\begin{table}
\caption{Stability Indices for $q_3(r)$ identification at different noise levels $h$.}

\begin{tabular}{r r r r r}
\hline

$k$ & $Iteration$ & $h=0.00$ & $h=0.01$ & $h=0.10$\\

\hline

\hline

1.00 & 1 & 0.564168 & 0.594314	 & 0.764340 \\
   & 2 &   0.024441  & 0.028558 & 0.081888 \\
   & 3 &   0.024441  & 0.014468 & 0.050755 \\
   & 4 &   0.024684  &  &  \\
   & 5 &   0.024684  &  &  \\
   & 6 &  0.005800   &  &  \\

\hline

2.00 & 1 & 0.684053 & 1.450148	 & 0.485783 \\
   & 2 &  0.423283   & 0.792431 & 0.078716 \\
   & 3 &  0.006291   & 0.457650 & 0.078716 \\
   & 4 &     & 0.023157 & 0.078716 \\
   & 5 &     &  & 0.078716 \\
   & 6 &     &  & 0.078716 \\
   & 7 &     &  & 0.078716 \\
   & 8 &     &  & 0.078716 \\
   & 9 &     &  & 0.078716 \\
   & 10 &     &  & 0.078716 \\
   & 11 &     &  & 0.078716 \\

\hline

2.50 & 1 & 0.126528 & 0.993192	 & 0.996519 \\
   & 2 &  0.013621   & 0.105537 & 0.855049 \\
   & 3 &     & 0.033694 & 0.849123 \\
   & 4 &     & 0.026811 & 0.079241 \\
\hline

3.00 & 1 & 0.962483 & 1.541714	 & 0.731315 \\
   & 2 &  0.222880   & 0.164744 & 0.731315 \\
   & 3 &  0.158809   & 0.021775 & 0.072009 \\
   & 4 &  0.021366   &  &  \\
   & 5 &  0.021366   &  &  \\
   & 6 &  0.001416   &  &  \\

\hline
4.00 & 1 & 1.714951 & 1.413549	 & 0.788434 \\
   & 2 &  0.033024   & 0.075503 & 0.024482 \\
   & 3 &  0.018250   & 0.029385 &  \\
   & 4 &     & 0.029421 &  \\
   & 5 &     & 0.029421 &  \\
   & 6 &     & 0.015946 &  \\

\hline

\end{tabular}
\end{table} 

For example, Table 4 shows that for $k=2.5,\ h=0.00$ the Stability Index has reached
the value $0.013621$ after 2 iteration. According to the Stopping
criterion for IRRS, the program has been terminated with the conclusion
that the identification was stable. In this case the  potential
identified by the program was

\[
p(r)=\begin{cases} 
-10.000024 & 0\leq r < 7.999994  \\
0.0 &  r \geq 7.999994
\end{cases}
\]

which is very close to the original potential

\[
q_3(r)=\begin{cases} 
-10.0 & 0\leq r < 8.0\\
0.0 &  r \geq 8.0
\end{cases}
\]

On the other hand, when the phase shifts of $q_3(r)$ were corrupted by a
$10\%$ noise ($k=2.5,\ h=0.10$), the program was terminated 
(according to the Stopping criterion) after 4
iterations with the Stability Index at $0.079241$. Since the Stability
Index is greater than the a priori chosen threshold of $\epsilon=0.02$
the conclusion is that the identification is unstable. A closer look
into this situation reveals that the values of the objective function
$\Phi(p_i),\ p_i\in S_{min}$ (there are 8 elements in 
$S_{min}$) are between $0.0992806$ and $0.100320$.
Since we chose $\beta=1.1$ the values are within the required $10\%$ of
each other. The actual potentials for which the normalized distance is equal to
the Stability Index $0.079241$ are
\[
p_1(r)=\begin{cases} 
-9.997164 & 0\leq r < 7.932678\\
-7.487082 & 7.932678 \leq r < 8.025500\\
0.0 &  r \geq 8.025500
\end{cases}
\]
and
\[
p_2(r)=\begin{cases} 
-9.999565 & 0\leq r < 7.987208\\
-1.236253 & 7.987208 \leq r < 8.102628\\
0.0 &  r \geq 8.102628
\end{cases}
\]
with $\Phi(p_1)=0.0992806$ and $\Phi(p_2)=0.0997561$.
One may conclude from this example that the threshold $\epsilon=0.02$ is too tight
and can be relaxed, if the above uncertainty is acceptable.


\section{Conclusions}
A novel numerical method for solving inverse scattering
problem with fixed-energy data is proposed.
The method contains a new important concept: the stability index
of the inversion problem. This index is a number, computed from the data,
which shows how stable the inversion is. If this index is small, then the
inversion provides a set of potentials which differ so little, that
practically one can represent this set by one potential. If this
index is larger than some threshold, then one concludes
that practically,
 with the given data, the inversion is unstable and the potential
cannot be identified uniquely from these data.
Inversion of the
fixed-energy phase shifts for several model potentials
is considered.
The results show practical efficiency of the proposed method.
The method is of general nature and is applicable to a very wide variety
of inverse problems.

\end{document}